# A Comparative Study of Association Rule Mining Algorithms on Grid and Cloud Platform

Sudhakar Singh[1], Rakhi Garg[2], P. K. Mishra[3]
Department of Computer Science - Faculty of Science[1,3], Mahila Maha Vidyalaya[2]
Banaras Hindu University[1,2,3]
Varanasi – 221005, UP, India[1,2,3]
sudhakarcsbhu@gmail.com[1]   rgarg@bhu.ac.in[2]   mishra@bhu.ac.in[3]

*Abstract*: Association rule mining is a time consuming process due to involving both data intensive and computation intensive nature. In order to mine large volume of data and to enhance the scalability and performance of existing sequential association rule mining algorithms, parallel and distributed algorithms are developed. These traditional parallel and distributed algorithms are based on homogeneous platform and are not lucrative for heterogeneous platform such as grid and cloud. This requires design of new algorithms which address the issues of good data set partition and distribution, load balancing strategy, optimization of communication and synchronization technique among processors in such heterogeneous system. Grid and cloud are the emerging platform for distributed data processing and various association rule mining algorithms have been proposed on such platforms. This survey article integrates the brief architectural aspect of distributed system, various recent approaches of grid based and cloud based association rule mining algorithms with comparative perception. We differentiate between approaches of association rule mining algorithms developed on these architectures on the basis of data locality, programming paradigm, fault tolerance, communication cost, partition and distribution of data sets. Although it is not complete in order to cover all algorithms, yet it can be very useful for the new researchers working in the direction of distributed association rule mining algorithms.

## I. Introduction

The explosive growth of data from various resources like business, scientific and social networks increases rapidly and is crossing petabyte scale. These huge raw data is not significant until converted to small, useful, precise information which is able to process by human brain for further decision making process. Knowledge Discovery comes into existence in order to extract useful information and knowledge from such large amount of data. Data mining is a major step in the process of knowledge discovery in database. Data mining refers to discovering or extracting hidden useful and interesting knowledge from large amount of data [15]. Association Rules Mining (ARM) is one of the most important functionality of data mining which finds a correlation or association among attributes or items of the database. The typical example of ARM is Market Basket Analysis that discovers which items are frequently purchased together by the customers. The business applications of ARM are decision making, cross marketing, marketing promotion. ARM is also applicable to bio-informatics and medical diagnosis [15].

The traditional ARM algorithms have sequential nature and do not provide scalability to huge data sets. To overcome this pitfall high performance ARM algorithms are proposed which are scalable and taking less response time [6]. Although both parallel and distributed algorithms improve the performance of traditional algorithms but differ in system architecture used by them. In parallel system ARM algorithm executes on either shared or distributed memory machine or clusters enabled with fast network whereas in distributed systems, ARM algorithm executes on loosely-coupled system like clusters having nodes connected by slow networks and geographically distributed [7]. It is possible to apply traditional algorithms in distributed environment by collecting distributed data at a single physical place but it will be practically inefficient since it involves high storage cost, communication cost and computational cost [11]. Distributed mining also becomes necessary if the data is inherently distributed and sensitive [11]. Traditional parallel and distributed ARM algorithms are designed on homogeneous platform and partition the database evenly across nodes but it does not fit in heterogeneous environment like grid and cloud [7].

Cluster computing provides the base for both grid and cloud computing that administrated and operated by a single organization or center. It is built of commodity computers of similar or identical nature that accommodate at the same physical location where each constituting machine has its own memory, disk and network interface [38]. Here, computers are connected by high speed LANs and are less costly than supercomputers [38]. Grid is cluster of geographically distributed heterogeneous machines [30], [38]. Since a grid is geographically distributed so wide area networks are used to connect the nodes. Such networks have less bandwidth and higher latency than as in clusters. The constituting machines in grid are not under the same administration [38]. On the other hand cloud computing not only impressively overlaps with grid computing, it also uses infrastructure of grid as its backbone. Grid infrastructure delivers services of storage and computational resources while cloud purely grounded on service oriented architecture delivers economy based services as more abstract resources [37]. All these distributed system are not exclusive but overlap to one another.

In this paper, we have reviewed the various grid based and cloud based ARM algorithms, differentiates between approaches for ARM algorithm to be developed on these architectures and focus on the important issues of each ARM algorithm discussed to help researchers working in this area. The rest of the paper is organized as follows: Section II introduces ARM technique and reviews various sequential and parallel algorithms. Section III contrasts the architectural details of grid and cloud system. Various grid and cloud based ARM approaches are compared in section IV. Finally in section V, we conclude the paper and focus on future works.

## II. Association Rule Minig

### A. Basic Concepts

The formal statement of the Association Rule as described in [17] is as: Let $I = \{i_1, i_2, ..., i_m\}$ be set of attributes called items. Let D be a database containing set of transactions. Each transaction T is a set of items called itemset such that $T \subseteq I$. An association rule is an implication of the form $X \rightarrow Y$ where $X, Y \subset I$ are itemsets and $X \cap Y = \emptyset$. The strength of Association Rule is measured in terms of support and confidence which are defined as [15], [16]:

$$\text{Support, } s(X \rightarrow Y) = \sigma(X \cup Y) / N$$

$$\text{Confidence, } c(X \rightarrow Y) = \sigma(X \cup Y) / \sigma(X)$$

Where $\sigma(X) = |\{t_i | X \subseteq t_i, t_i \in T\}|$, denotes the support count or occurrence frequency of an itemset 'X'. The support, s is the percentage of transaction in D that contain $X \cup Y$ i.e. both X and Y and the confidence, c is the percentage of transaction in D containing X that also contains Y. Support is generally used to remove the uninteresting rules while confidence measures the reliability of the outcome obtained from the rule [16]. Association rule mining is a two-step process [15]: (i) Frequent Itemset Generation and (ii) Strong Association Rule Generation. Frequent itemsets are those that satisfy the user defined minimum support threshold min_sup i.e. itmeset having support greater than or equal to the min_sup [15]. Strong association rules are the high confidence rule generated from frequent itemsets i.e. rules having confidence greater than or equal to the minimum confidence threshold min_conf [15].

### B. Association Rule Mining Algorithms

Majority of the research has been given attention to efficient algorithm for frequent itemsets generation which is more expensive due to scanning whole database iteratively. R. Agrawal and R. Srikant [19] proposed the most popular Apriori Algorihm to mine the frequent itemsets using candidate itemsets generation. Many techniques have been proposed to improve the efficiency of the Apriori algorithm. These are Hash-based technique, Transaction reduction, Partitioning, Sampling and Dynamic itemset counting (DIC) [16]. Many sequential and parallel algorithms have been proposed for ARM in order to enhance scalability and performance. Some of these algorithms are candidate based (e.g. Apriori like) while some are without candidate generation (e.g. FP-growth) [21] and some others are based on lattice traversal scheme (e.g. Eclat, MaxEclat, Clique, MaxClique) [20]. Parallel ARM algorithms are categorized on the basis of shared, distributed and hierarchical memory systems, data distribution strategy, type of parallelism, load balancing technique e.g. static or dynamic used (e.g. CD, DD, PDM, IDD, CandD, HD, ParEclat, ParMaxEclat, ParClique, ParMaxClique etc.) [6], [12].

## III. Distributed System Architecture

In this section, we mainly focus on the architectural details of grid and cloud distributed computing systems.

The various protocols and services provided by grid are divided into five layers [44]. Cloud systems share various same attributes with clusters and grids system. Fig. 1 describes the four-layer architecture of cloud in contrast to five-layer grid architecture [37].

The layers in grid from lower to upper are fabric, connectivity, resource, collective and application while in cloud are fabric, unified resource, platform and application layer. Two layers named **Fabric** and **Application** are common in these two architectures. The functionality of these two layers in grid and cloud is shown in Table I [44], [37].

**Figure 1 Grid Protocol Architecture vs. Cloud Architecture [44], [37]**

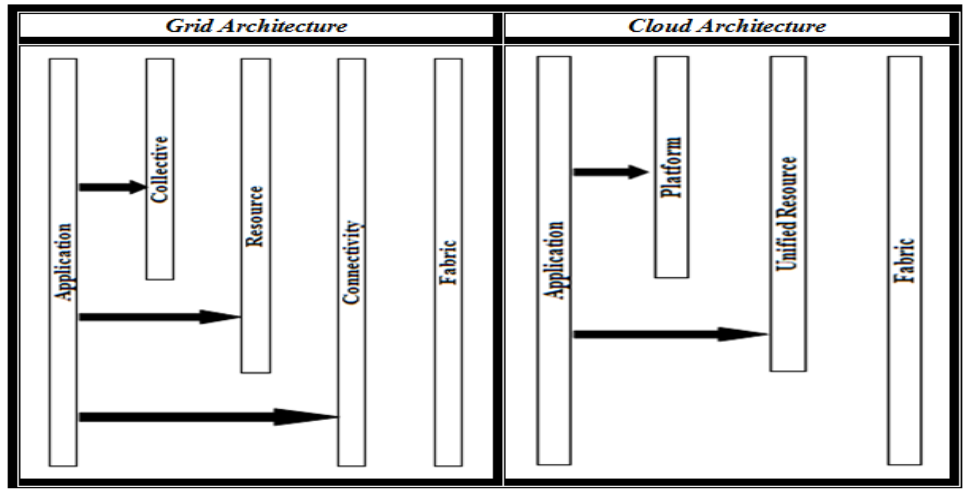

**Table I Common Layers in Grid and Cloud**

| Layers | Grid Architecture | Cloud Architecture |
|---|---|---|
| Fabric Layer | Interfaces to shared resources such as computational resources, storage and network resources, specialized storage resources: code repositories and catalogs. | Covers the resources at hardware level such computational resources, storage and network resources. |
| Application Layer | Comprises the user's applications that operate in virtual organizations and uses the layers above fabric layer to perform grid operations. | Simply contains the applications to be run in the clouds. |

In grid, the **connectivity layer** provides protocols for core communication and authentication to enable data exchange easily and securely between fabric layer resources [44]. The **resource layer** defines protocols for resource sharing. It builds on connectivity layer for controlled access of resources. It obtains information about resources and negotiates the access of shared resources by two classes, information protocols and management protocols respectively [44]. The **collective layer** addresses interactions among multiple resources. It integrates multiple resources together, e.g. it can build virtual data source by combining data sources from multiple sites and get the result from computations on multiple sites [18].

In cloud, the **unified resource layer** contains abstracted or encapsulated resources which can be accessed as integrated resources by upper layer and end users, e.g. virtual computer or cluster, logical file system, database system etc [37]. The **platform layer** provides development or deployment platform such as web hosting environment, scheduling service etc. by adding a collection of specialized tools, middleware and also services above the unified resources [37].

The connectivity, resource and collective layers in grid architecture are termed as **middleware layer**. There is various middleware software to implement middleware layer and Globus Toolkit is the most popular implementation. The **globus toolkit** [35] is an open source software and most widely used toolkit for building service oriented grid infrastructures and applications. Globus Toolkit is used to develop the application whose objective is to combine distributed resources which may be workstations, storage, data sets, software networks, sensors etc [18].

There are various organizations and companies facilitating cloud services. Google's cloud hardware is based on the cluster of a large number of commodity computers rather than very high performance computers. Google's design pattern is based on Google File system (GFS) [41] as storage system and MapReduce [26], [40] framework as the computational model. Amazon's web services Elastic Compute Cloud (EC2) [1] and Simple Storage Service (S3) [2] facilitate on demand compute capacity in cloud and interface to store and access any sized data from anywhere on the web in a secure manner. Similar to Amazon, Microsoft launched its cloud based Azure Services Platform [3] to provides environments in which one can develop, host and manage cloud applications.

### IV. ARM on Grid and Cloud Architecture

It has been observed that there are differences in grid and cloud architecture, so there might be some differences in the approach of developing ARM algorithms on these environments also. In this section, we analyze the ARM algorithms on these two environments and compare their approaches on some common issues and parameters.

### *A.   ARM Algorithms for  Grid Computing Environment*

Association Rule Mining in grid environment causes workload imbalances, which has two reasons: (i) workload imbalance due to heterogeneous nature of grid environment and (ii) correlation of itemsets cannot be predicted before execution of ARM algorithms [25]. Various grid data mining projects such as Discovery Net (D-NET) [13], Grid Miner [31]--[34], Knowledge Grid [14] only integrate and deploy classical algorithms on grid but there is not any efficient algorithm suitable for grid architecture [25]. To design an efficient grid based algorithms we have to focus on appropriate data partitioning and distribution approach, load balancing strategy, optimization of synchronization and communication. Various ARM algorithms for grid environment have proposed that focus on these issues. Some of them are:

*1) DisDaMin Project (DIStributed DAta MINing)*- proposed by V. Fiolet et al. [28] that uses intelligent data distribution for ARM which depends on data fragmentation using clustering methods. The intelligent fragmentation method minimizes the communication by computing fragments on each node. After intelligent fragmentation has been completed, a distributed algorithm DICCoop inspired by DIC is used. It uses a federator to merge local result of fragments to make global results.

*2) Grid-based Distributed Max-Miner (GridDMM)-* proposed by C. Luo et al. [29] that mines the maximal frequent itemsets on a Data Grid System which is based on Globus Toolkit. Database is evenely distributed across nodes of Data Grid. GridDMM consists of two phases: local mining and global mining. During local mining phase, each node uses sequential Max-Miner algorithm on its local database to find local maximal frequent itemsets [9]. After that set of maximal candidate itemsets are formed for top-down search in global mining phase. A prefix tree data structure is used to store and count the global candidate itemsets.

*3) Distributed Algorithm to Generate Frequent Itemsets on Heterogeneous Clusters and Grid Environment-* proposed by Lamine M. Aouad et al. [22] that takes into account the inherent dynamic nature associated to the candidate set generation and memory constraints of each node and uses dynamic workload management with block-based partitioning. The data set is divided into m partitions where m is the number of nodes. Each partition is divided into blocks and the size of each block depends on available memory in corresponding node, number of items, average transaction width and support threshold. The algoirthm consists of two phases. The first phase is the block-based apriori generation and workload management and the second phase is the remote support counts computation. To balance workload, the idle nodes select a donor processor from overloaded processors. The selected node should be of smaller block size and largest remote number of blocks. The objective of this strategy is to avoid the overhead of re-partitioning larger block into smaller blocks if the larger blocks do not fit in destination main memory.

*4) Heuristic Data Distribution Scheme (HDDS)-* proposed by Chao-Tung Yang et al. [23] for grid based data mining application. They formulate the data distribution problem by linear programming and proposed a heuristic algorithm to solve it. The algorithm is based on master/slave model. The MASTER module divide the total dataset among slaves based on the performance ratio (PR) of each slave and send the data to slaves in decreasing order of bandwidth of link to slaves. The performance ratio is determined by considering CPU speed, network bandwidth and memory capacity. The SLAVE module mines the local dataset and sends the result to the master. Master integrates results from the slave nodes.

*5) Hybrid Load Balancing Strategy-* proposed by R. Tlili and Y. Slimani which is combination of two parts: static and dynamic. The static part is proposed as a *Novel Data Partitioning Approach* [25]. In dynamic part, they propsed a *Hierarichal Model of Grid and Dynamic Load Balancing Strategy* [7], [10], [24].

The hierarchical grid model consists of T sites heterogeneous in computing and storage power. The set of sites is denoted as $G = (S_1, S_2, …, S_T)$, where each site is defined as $S_i = (M_i, Coord(S_i), L_i)$, where $M_i$ represents total number of cluster in site $S_i$. $Coord(S_i)$ is the coordinator of $S_i$ which balance the workload by moving work from overloaded cluster to lightly loaded cluster inside the same site called as intra-site or to remote site called as inter-sites. $L_i$ denotes the computational load at $S_i$. Each cluster is denoted as $cl_{ij} = (N_{ij}, Coord(cl_{ij}), L_{ij}, \omega_{ij})$, where $N_{ij}$ is the number of nodes in cluster $cl_{ij}$, $Coord(cl_{ij})$ is the coordinator of $cl_{ij}$ which dynamically distribute the candidates among its nodes, $L_{ij}$ and $\omega_{ij}$ are computational load and processing time of cluster $cl_{ij}$ respectively. The database is partitioned by novel data partitioning approach discussed below. To balance the workload first migration of work takes place between nodes of the same cluster. If this is not sufficient then migration takes place between clusters of same site. And finally if there is still workload imbalance then migration of work between sites takes place [24]. Whereas the novel data partitioning approach is a preprocessing phase takes place before dynamic load balancing phase and categorized into three variants: Fair Partitioning, Homogeneous-content based Partitioning and Content-capacity-based Partitioning. Fair partitioning variant partitions the global database into equal sized partitions and allocate them to each site. Homogeneous content partitioning ensures the same work load production during execution. Content capacity partitioning is an improvement over homogeneous content partitioning that considers heterogeneity of the grid environment [25].

ARM algorithms on Grid Architecture discussed above are based on the three strategies which are data distribution, dynamic load balancing and hybrid approach i.e. data distribution plus dynamic load balancing. The classification of these algorithms on this basis is described in Table II.

**Table II Categorization of ARM Algorithms for Grid Architecture on the Basis of Data Distribution & Load Balancing**

| Data Distribution | Dynamic Load Balancing | Hybrid Approach |
|---|---|---|
| DisDaMin Project, GridDMM | HDDS | Distributed Algorithm to Generate Frequent Itemsets on Heterogeneous Clusters and Grid Environment, Hybrid Load Balancing Strategy |

## B. ARM Algorithms on Cloud Computing Environment using MapReduce

It is assumed that all the nodes in grid are fail-safe but as the number of nodes increases, the chance of failure of nodes increases [36]. The nodes failure cannot be ignored since it introduces incomplete data and might causes multiple re-execution of the jobs. Whereas adding more nodes increases the scalability but failure of nodes in grid limits it [36]. To overcome this pitfall, Google has developed MapReduce [26], [40] framework and programming model, which is a computational model used in cloud computing and uses the Google File System (GFS) [41] as storage system for reading and writing files. Hadoop [27], an open source project of Apache has implemented Google's cloud approach and provides Hadoop Distributed File System (HDFS) as data storage system and Hadoop MapReduce as data processing system. The MapReduce model is based on two functions. The map function takes input a list of (key1, value1) pairs and output intermediate (key2, value2) pairs. The intermediate key/value pairs are grouped by equal key basis as (key2, list (value2)) and passed to the combiner who makes local sum for each key2. The reduce function operated on the output of combiner and output the global result. Combiner reduces the communication by directly not providing output of map function to reduce function [39]. Some of ARM algorithms which have been designed over MapReduce framework are shown in Figure 2.

**Figure 2   ARM Algorithms on MapReduce Framework.**

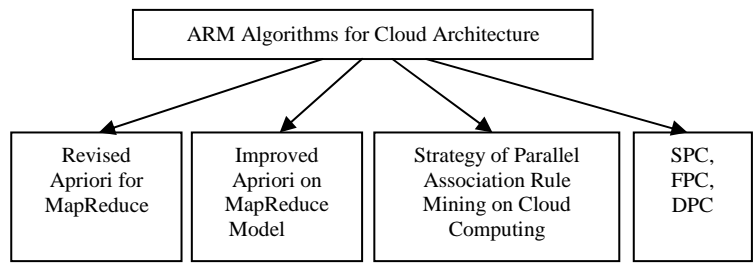

*1) Revised Apriori for MapReduce-* proposed by Juan Li et al. [8] that simply revise the Apriori for MapReduce and implemented it on Hadoop framework. Datasets is splited by the "InputFormat" method of Hadoop which by default breaks up a file into 64 MB chunks and also can be configured. The splited chunk is assigned to mapper. Mapper reads one line at a time and output the (key, value) pairs, where key is assigned to each item and value is 1 in order to find frequent-1 itemsets. Combiner takes (key, value) pairs and produce the list of value for each key and sum up the list of value of each key. The output of the combiner function is taken as input for reduce task which sum up the count of values associated with each key. Items having value less than the minimum support are pruned and remaining items termed as frequent itemsets are written to output file. Candidate itemsets are generated from frequent itemsets of previous iteration. Candidate itemset is considered as a key if it is found in transaction set assigned to mapper and assigned value as 1. Further procedure is similar to frequent-1 itemset generation. It is repeated until output file from the previous iteration is not found. There is sub-folder of each iteration containing the frequent itemsets with support count which are further used to generate association rules. The algorithm was deployed on Amazon EC2 and S3 [1], [2].

*2) Improved Apriori Algorithms on MapReduce Model-* proposed by Xin Yue Yang et al. [4] that is similar to the above aproach. Map ( ) and Reduce ( ) are used in Apriori algorithm to find the subset of the candidate itemsets and to find the frequent itemsets respectively. This approach simply converts the apriori algorithm to MapReduce model. Map function operate on data segments and outputs (key, value) pairs for every records assigned to it. All the pairs are grouped on the basis of same item and passed the list of values to reduce function to produce a count for the candidate itemsets. During every scan, local candidate itemsets are generated by map function and reduce function generates global count by aggregating local counts.

*3) Strategy of Parallel Association Rule Mining on Cloud Computing-* proposed by L. Li and M. Zhang [5] in which database is divided into n data subsets by MapReduce library and sent to m Map tasks executing on m nodes. The n data subsets are formatted as (key, value) pairs, where key is Tid, transaction identifier and value is list of corresponding transaction in database. The Map function processes the (Tid, list) record and generate local candidate itemsets as (itemset, 1). The combiner function combines the sorted output i.e. (itemset, list (1)) of Map function and generate (itemset, sup), where sup is support count in subset of database. A partition function *hash (key) mod R* used to divide output of combiner to R different partitions assigned to R Reduce tasks. Reduce function find the actual support count in the whole database. A *data set distribution method* [5] is also proposed which uses 16 MB as distribute unit for better load balance. Dataset is divided into n non intersecting segments where n ≥ 4m and m is the number of nodes executing Map tasks whereas the number of distribution data set for each node is determined on the basis of computing capability of the node.

*4) SPC, FPC, DPC-* proposed by Ming-Yen Lin et al. [36] which are three variants of apriori algorithms over MapReduce framework named as *Sigle Pass Counting (SPC), Fixed Passes Combined-counting (FPC)* and *Dynamic Passes Combined-counting (DPC).* SPC, the most basic is simply an implementation of the apriori algorithm in MapReduce framework like algorithms discussed above. Due to multi-pass nature of apriori, MapReduce implementation involves multiple map-reduce phases. The map function must have to wait to finish the previous phase's reduce functions before starting next phase. All the nodes having completed their reduce task have to wait other nodes not completing their reduce tasks. This leads to the scheduling and waiting overheads. FPC and DPC implement apriori in MapReduce framework more efficiently by minimizing scheduling invocation, maximizing node utilization and workloads balance. Table III distinguishes the characteristics of these algorithms. In FPC map function may overload if merging of passes generates large number of candidates. DPC overcomes this problem and dynamically merges consecutive passes taking into account the workload of workers. The innovative ideas of FPC and DPC are same except that former statically combines the candidates and latter combines on basis of the candidate threshold calculated dynamically.

**Table III SPC, FPC and DPC**

| SPC | FPC | DPC |
|---|---|---|
| Each map-reduce phase operate on single pass of Apriori. | Merges consecutive fixed number of passes into a single map-reduce phase. | Combined passes into a map-reduce phase dynamically to maximize the node utilization. |
| Generates frequent k-itemsets in a map-reduce phase at k-th scanning of database. | Generates frequent k-, (k+1)-, …, and (k+m)-itemsets (m=2) in only a single map-reduce phase. | Generates and collects various length candidate itemsets such that total number of candidates is not crossing a pre-calculated value, candidate threshold. |

*C. Issues and Parameters for Grid and Cloud based Approach*

The objective of both Grid and Cloud are to reduce computational cost, enhance reliability and flexibility. Most of the issues are same that may differ in technical details and technologies used. In general, Grid is a federated system and cloud is commercialized system [37]. Keeping into account the characteristics of distributed ARM algorithms; Table IV contrasts the approaches in these two distributed computing environments on the basis of some common issues and parameters.

**Table IV Cloud Based Vs. Grid Based ARM**

| Issues and Parameters | Cloud based Approach | Grid based Approach |
|---|---|---|
| Data Storage and Data Locality | Data Storage System: GFS, HDFS; Efficient data locality. | Shared file system e.g. NFS, GPFS, PVFS, Luster; Extra effort is required for data locality. |
| Programming Paradigm | MapReduce | MPI, MPICH-G2 |
| Fault Tolerance | Highly Fault Tolerant | Independent tasks may ensure Fault Tolerance. |
| Communication Cost | Low | High |
| Partition and Distribution of Data Sets | Done by MapReduce Library. Heuristics can also be used. | Done by Heuristics. |

Some of the issues are:

*1) Data Storage and Data Locality:* In order to enhance scalability we have to distribute data across many nodes and perform computation on data segments incurring minimum communication overhead. GFS and HDFS partition the data into many chunks and replicate them to exploit efficient data locality. Grid use shared file system e.g. NGS, GPFS, PVFS, Luster which is not suitable to exploit data locality [37].

*2) Programming Paradigm:* Message Passing Interface (MPI) is the most widely used programming paradigm to develop parallel applications on distributed memory architecture [43]. A set of processes communicate to one another by sending and receiving messages using point to point and collective communication routines in MPI. MPICH-G2 is a Grid enabled implementation of the Message Passing Interface [42]. MPICH-G2 is redesign and reimplementation of the MPICH-G system to significantly increase the performance. MapReduce is a simplified programming model used in cloud computing for processing large volume of data sets [26], [40]. The advantage of MapReduce is that it allows focusing only on the computation and not on the parallelization [40]. The run time system automatically partitions the input data, schedules the execution of program over machines, handles machine failures and manages inter-machine communications.

*3) Fault Tolerance:* As the number of nodes in grid increases there is a potential chance of node failures. MapReduce provides scalability with robust fault tolerance. MapReduce can continue working despite of 1.2 average worker failures per job run at Google [39], [40]. In grid we have to partition the problems into independent tasks which ensure fault tolerance more than those having complex communication.

*4) Communication Cost:* MapReduce framework is based on shared-nothing architecture. Each slave node can communicate only with master node and prohibited to communicate with other slave nodes which lead to low communication cost. In grid environment any node can communicate with other nodes incurring high communication cost.

*5) Partition and Distribution of Data Sets:* MapReduce library splits the input file into chunks which can be between 16 MB to 64 MB [5]. Chunks are replicated across many nodes to face node failures. Grid uses various heuristics to partition the data sets among nodes. A performance based approach may be used to distribute the datasets across working nodes that can be defined by attributes like CPU speed, memory and network bandwidth etc.

## V. Conclusion

Various ARM algorithms for grid as well as cloud computing have been developed. Among algorithms developed for grid environment, DisDaMin Project and GridDMM follow only data distribution while HDDS follows only dynamic load balancing. An efficient algorithm must follow hybrid approach i.e. data distribution and dynamic load balancing. There are two algorithms which follow hybrid approach. One is Frequent Itemsets Generation on Heterogeneous Clusters and Grid Environment and second one is Hybrid Load Balancing Strategy. Among algorithms developed for cloud environment, first two simply implement Apriori on MapReduce framework while third one also adds data set distribution method. SPC, FPC and DPC are the three variant of apriori on MapReduce in which DPC is the most efficient and maximize the node utilization. Running ARM algorithms on grid is more challenging than on cloud using MapReduce. Grid is a federated system while cloud is economy based and it is easy to access cloud resources on demand. Both the grid and cloud based approaches are facing some common issues which are data locality, handling fault tolerance, communication minimization and partition and distribution of data sets and more research has to be done on these issues to develop efficient algorithms.